\documentclass[reprint,
					aps,
					pra,
					english,
					footinbib,
					10pt,
					superscriptaddress
]{revtex4-1}

\usepackage[english]{babel}

\usepackage[utf8]{inputenc}

\usepackage[T1]{fontenc}

\usepackage[intlimits]{amsmath}

\usepackage{amssymb}

\usepackage{mathtools}

\usepackage{dsfont}

\usepackage{siunitx}

\usepackage[final,colorlinks=true,linkcolor=blue,citecolor=blue]{hyperref}

\newcommand*{\defeq}{\mathrel{\vcenter{\baselineskip0.5ex \lineskiplimit0pt
	\hbox{\scriptsize.}\hbox{\scriptsize.}}}%
=}

\DeclarePairedDelimiterX\braket[2]{\langle}{\rangle}{#1\,\delimsize\vert\,\mathopen{}#2}
\DeclarePairedDelimiterX\matrixel[3]{\langle}{\rangle}{#1\,\delimsize\vert\,\mathopen{}#2\,\delimsize\vert\,\mathopen{}#3}

\DeclareMathOperator\perm{perm}

\DeclarePairedDelimiter\ceil{\lceil}{\rceil}

\usepackage{verbatim}

\begin{document}
\title{Boson sampling with random numbers of photons}
\author{Vincenzo Tamma}
\email{vincenzo.tamma@port.ac.uk}
\affiliation{School of Mathematics and Physics, University of Portsmouth, Portsmouth PO1 3QL, United Kingdom}
\affiliation{Institute of Cosmology \& Gravitation, University of Portsmouth, Portsmouth PO1 3FX, UK}
\affiliation{Institut f\"{u}r Quantenphysik and Center for Integrated Quantum Science and Technology (IQ\textsuperscript{ST}), Universität Ulm, D-89069 Ulm, Germany}
\author{Simon Laibacher}
\affiliation{Institut f\"{u}r Quantenphysik and Center for Integrated Quantum Science and Technology (IQ\textsuperscript{ST}), Universität Ulm, D-89069 Ulm, Germany}


\begin{abstract}
Multiphoton interference is at the very heart of quantum foundations and applications in quantum sensing and information processing.
	In particular, boson sampling experiments have the potential to demonstrate quantum computational supremacy while only relying on multiphoton interference in linear optical interferometers.
	However, even when photonic losses are negligible, scalable experiments are challenged by the rapid decrease of the probability of success of current schemes with probabilistic sources for a large number of single photons in each experimental sample.
	Remarkably, we show a novel boson sampling scheme where the probability of success increases instead of decreasing with the number of input photons  eventually approaching a unit value even with non deterministic sources.
	This is achieved by sampling at the same time in the number of occupied input ports and the number of input photons per port, differently form previous schemes where the number of occupied ports is fixed at each experimental run.
	Therefore, these results provide a new exciting route toward future demonstrations of quantum computational supremacy with scalable experimental resources.
\end{abstract}

\maketitle

Boson sampling \cite{Aaronson2011_computationalcomplexitylinear} has recently triggered the interest of both the quantum optics and computer science communities because of its potential to outperform classical computers, while relying only on the interference of $N$ single photons in linear optical networks \cite{Tamma2015_MultibosonCorrelationInterferometry}.
Indeed, it is simply defined as the task of sampling from the probability distribution where $N$ photons are found at the output of a random interferometer with $M \sim N^2$ ports.
This has triggered several experimental demonstrations with relatively small numbers of photons \cite{Broome2013_Photonicbosonsampling,Crespi2013_Integratedmultimodeinterferometers,
Tillmann2013_Experimentalbosonsampling,Spring2013_Bosonsamplingphotonic,
Bentivegna2015_Experimentalscattershotboson,Wang2017_Highefficiencymultiphotonboson,Pan2019BosonSampling}.
The computational hardness of boson sampling has been also demonstrated in the case of nonidentical photons by introducing the problem of multi-boson correlation sampling, where additional sampling in the phtonic inner modes, including time, frequency and polarization, allows to take advantage of the full information encoded within the photonic mode structure \cite{Laibacher2015_PhysicsComputationalComplexity,Tamma2015_Multibosoncorrelationsampling, Laibacher2018,Tamma2014_Samplingbosonicqubits}. Experimental implementations of such a  problem where carried out with photons either generated at different time or of different colors \cite{Wang2018ExpMBCS,Orre2019ExpMBCS}.

An experimental race to develop boson sampling demonstrations for larger and larger photon numbers $N$  is ongoing to be able to beat current classical algorithms with state of the art supercomputers  \cite{Pan2019BosonSampling}.  Indeed, the development of faster and faster classical algorithms require a number $N > 50$ of photons  to make the problem classically intractable \cite{Neville2017_Classicalbosonsampling}. More recently, in Ref. \cite{Dalzell2020howmanyqubitsare} it was claimed that a sampling circuit with 98 photons  and 500 linear optical elements  would be necessary to exceed a one century computation time with state-of-the-art supercomputers.

Unfortunately at the increase of $N$ the probability of success of standard boson sampling schemes  decreases exponentially given the non deterministic nature of photonic sources.
A first success in overcoming this drawback was achieved with the introduction of Scattershot Boson Sampling (SBS) \cite{Lund2014_BosonSamplingGaussian, Aaronson2013_ScattershotBosonSamplingnew, Bentivegna2015_Experimentalscattershotboson}.
Here, $M \sim N^2$ spontaneous parametric down conversion (SPDC) sources, one at each interferometric input port, are employed to generate in postselection $N$ identical single photons in a heralded, but random set of spatial channels.
Unfortunately, the scaling of the success probability still decreases with $N$, namely as $1/\sqrt{N}$, and the requirement of $M\sim N^2$ challenge experimental realizations at increasing values of $N$.

The more recent Gaussian boson sampling problem based on the use of non heralded sources of squeezed  states relies more generally on the computational hardness of matrix Hafnians and led to interesting still open questions on the computational hardness of multiphoton interference with squeezed light \cite{Hamilton2017GBS,Hamilton2019GBS}. However,  also this problems suffers from the same scaling  as $1/\sqrt{N}$ of the probability of detecting single photons in $N$ separate channels \cite{Hamilton2017GBS,Hamilton2019GBS}.

Therefore, all the current boson sampling schemes are still challenged by the  inability of successfully generating a multi-photon sample with a probability which is not highly suppressed at increasing number of photons. Indeed, even a decreasing in the success probability as $1/\sqrt{N}$  would lead to small probability values for large enough values of $N$ (at the moment  $N \sim 100$ based on current supercomputer computational power \cite{Dalzell2020howmanyqubitsare}). 

Indeed, toward obtaining the desired quantum computational supremacy in an experimental scalable manner there is  still an important question  to be addressed:
Is it possible to implement boson sampling schemes with a number of sources only linear in $N$ able to circumvent completely the probabilistic nature of photonic sources and generate photonic samples  with  probability which does not scale down with  $N$?
Such a question is intimately connected with a more fundamental one:
Is boson sampling still hard if we include scenarios in which the total number of input photons can randomly vary between one experimental sample and another to  increase the number of successful sampling events?


In this letter, we demonstrate the quantum computational hardness  of multiphoton interference in quantum optics linear networks beyond any classical capabilities even for random values of the number of photons injected in the interferometer. Remarkably, we show how, by using only a linear number of sources, at the increasing of $N$ the success probability of generating  a sample does not decrease any more but increases and even approach a unit value. Indeed, this is possible by additionally sampling over the number of photons per input port without  disregarding  events where photons are injected in more than $N$ channels. Furthermore, we demonstrate how this can be achieved either in the case of single-photon or multi-photon emissions. The proposed boson sampling schemes may also bring advantages to experimental scenarios where losses cannot be neglected given that samples with a total number of photons larger than $N$ are also likely to be generated.
\begin{figure}[!ht]
   \centering
   \includegraphics[width=.5\textwidth]{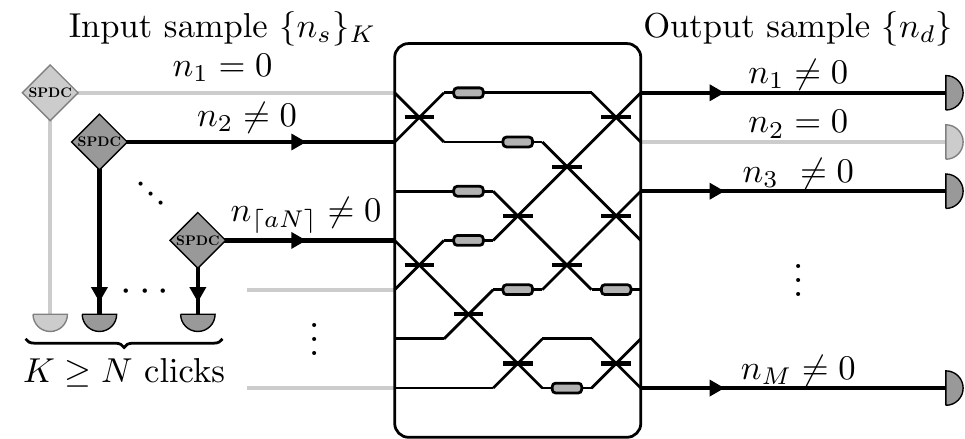}
\caption{Optical setup for Random Number Boson Sampling (RNBS), with $ \ceil  {a N} $ heralded SPDC sources, with $a>1$, placed  at the  input channels $s=1,..., \ceil  {a N}$  of a random $M$-channel linear interferometer ($\ceil{ }$ denotes the ceiling function). Each experimental run leads to the random selection of: (1)  a heralded input sample $\{ n_s \}_{K} $ of occupation numbers $n_s$, which are different from zero at a random set of $K \geq N$ input channels; (2) an output sample $\{ n_d \}$ of the detected photon numbers $n_d$ at each detector $d=1,...,M$, where at least $N$ detectors click. We have shown that such a scheme is computationally hard to simulate classically either in the scenario of single photon emissions ($n_s = 0,1$) or when multi-photon emissions are also likely to occur.}
	\label{setup}
\end{figure}

\textit{Random Number Boson Sampling (RNBS) with single-photon emission sources: sampling over a random number $K\geq N$ of input ports fed with single photons}.

We consider  the setup depicted in Fig. \ref{setup}  where a linear number $ \ceil  {a N} $ of heralded SPDC sources ($a>1$ and $\ceil{ }$ denoting the ceiling function) are connected to
the input ports $s = 1,...,\ceil  {a N}$ of a $M$-port linear network and the remaining
$M - \ceil  {a N}$ input ports are always fed with the vacuum state. We assume for simplicity at the moment that multi-photon emissions
of the sources are negligible. We consider all the samples where any random number $ K\in \{N, N+1,,...,\ceil  {a N}\} $ of single photons can be injected into the network at each experimental run and only events where $K < N$ are neglected.
We show that the corresponding sampling problem cannot be simulated classically in a time which is polynomial in the lower bound $N$ on the number of occupied input channels.

All the input samples  are defined by all the possible experimental sets of input occupation numbers  $\{ n_s \}_{K} $, with $s=1,...,\ceil  {a N}$, $n_s =0,1$, where a random number $K$  of  ports , with $N\leq K\leq \ceil  {a N}$,  are fed with single photons, while the remaining ones with the vacuum. This is substantially different from SBS where the number $K$ of occupied ports is fixed at each experimental run, namely $K=N$. Indeed, here,  we  do not neglect but take advantage of all additional events where more than $N$ sources generate single photons at the interferometer input channels. 
Furthermore, for each given  input sample $\{ n_s \}_{K} $ a sampling process also occurs at the interferometer output over all the possible set $\{ n_d \}$  of numbers $n_d$ of detected photons at each of the output ports $d=1,...,M$.     For simplicity, we can address  the case where  photon bunching is negligible, i.e. $n_d=0,1$, although as we will see later the only thing that matters in the computational hardness of the problem is that at least $N$ detector clicks for each sample.


We consider  a random $M \times M$ unitary matrix $U =\{U_{d,i}\}$ associated with a random  interferometer with  output ports $d=1,...,M$ and $\ceil  {a N}$ sources placed in  the subset of ports $s=1,...,\ceil  {a N}$ of the input ports $i=1,...,M$. At each experimental run, a random sample $\{ n_s ; n_d \}_{K}$ is generated where  $N\leq K\leq \ceil  {a N}$  input ports are occupied with single photons.  We will refer to the task of sampling from this probability distribution as random number boson sampling (RNBS) with single-photon emission sources.
Indeed, the probability of generating a given sample  $\{ n_s ; n_d \}_{K}$ relies on the  interferometer transition amplitudes $U_{\bar{d},\bar{s}}$ corresponding to an input channel $s=\bar{s}$ occupied by one photon ($n_{\bar{s}} =1$)  which is than detected by a detector $d=\bar{d}$ ($n_{\bar{d}} =1$). All non occupied input  channels where no photons are generated ($n_s=0$) and no sources are  present  and all the output channels where no photons are detected ($n_d =0$) are not involved into the interferometric evolution. Therefore,   it is useful to introduce for each sample $\{ n_s ; n_d \}_{K}$ a corresponding $K \times K$ matrix $U^{\{ n_s ; n_d \}_{K}}$ obtained by considering the interferometric matrix $U$ with only the columns $s=1,...,\ceil  {a N}$ associated with the input channels with a source and by taking each of these columns  $n_s$ times and each of the $M$ rows $n_d$ times. This leads to the probability
\begin{eqnarray}\label{BSprobSinglePhotons}
P^{\{ n_s ; n_d \}_{K}} \propto \mid perm \, U^{\{ n_s ; n_d \}_{K}} \mid ^2.
\end{eqnarray}
of generating any given sample  $\{ n_s ; n_d \}_{K}$, where we used the definition  of the permanent of a matrix $\mathcal{A}$  as $	\perm \mathcal{A} \defeq \sum_{\sigma\in\Sigma_N} \prod_{i=1}^{N} \mathcal{A}_{i\sigma(i)}$,

We emphasize that the matrices $U^{\{ n_s ; n_d \}_{K}}$ are for all samples bounded in size from below by $N$  and, as in standard boson sampling and SBS, their elements are all  i.i.d. Gaussian. Consequently, the permanents of such matrices  of rank $K\geq N$ are at least as hard to compute as in standard boson sampling and SBS where the matrix rank is $K=N$ for each sample. Therefore, approximate RNBS with single-photon emission sources is at least as computationally hard as boson sampling and SBS \cite{Barvinok1996Rank}. Indeed,  RNBS can be interpreted as sampling over different SBS configurations  each associated with a randomly picked number $K\in [N,\ceil  {a N}]$  of ports fed with single photons.

\textit{Generalised RNBS: sampling over a random number of input photons at each of the  $K$ occupied ports, with random values of $K \geq N$}.

So far we have considered the case of probabilistic single photon sources where the emission of more than one photon per source is negligible.  However, one may ask if it is possible to further enhance the scalability of boson sampling schemes by considering sources for which  sampling events based on multi-photon emissions can be also relevant  leading to a random total number  $N_{tot} = \sum_s n_s$ of input photons per sample. Such a value $N_{tot}$, differently from the previous scenario, can be larger than the number of occupied ports $K$  but it is still of the order of $N$ given that the $ \ceil{a N}$ input sources are independent of each other.

This leads to another fundamental question : is it still classically hard to solve the approximate sampling problem, when sampling over all the possible occupation numbers $n_s$ of input photons at each channel $s=1,...,\ceil{a N}$, by including events where $n_s > 1$ and not just $n_s =0,1$ ?

Remarkably, it is easy to show that, in this case, all the sample probabilities depend on permanents of  matrices $U^{\{ n_s ; n_d \}_{K}}$  as defined before Eq.  (\ref{BSprobSinglePhotons}). However,  now these matrices are not simply obtained by considering the $K$ columns associated with the occupied input channels but also by repeating each of these columns a number of times given by the corresponding occupation number $n_s$ which can be now larger than one. This leads to a total number $N_{tot} \geq K$ of columns, with still a random number $K\geq N$ of independent columns  associated with the $K$  occupied input channels. If we neglect then for simplicity bunching events,  a given output samples $\{n_d\}$ correspond to take only the relevant $N_{tot}$  rows associated with output occupation numbers $n_d =1$, and disregarding the ones with $n_d =0$. Therefore,  the matrices $U^{\{ n_s ; n_d \}_{K}}$, of random dimension $N_{tot} \geq K$,  contains $N_{tot} \times K \geq N^2$ entries which are i.i.d.\ Gaussian variables, with respect to standard boson sampling and Gaussian boson sampling where this number is fixed to be $N^2$ for each sample \cite{Aaronson2011_computationalcomplexitylinear}.
Therefore, analogously to the case before, the rank of  the matrices $U^{\{ n_s ; n_d \}_{K}}$ is always larger or equal to the fixed rank $N$ in current boson sampling schemes where a fixed total number $N_{tot} =N$ of photons are detected at each sample. This implies that also approximate boson sampling with a random number of input photons per port with $K \geq N$ occupied ports is at least as hard as standard boson sampling and SBS \cite{Barvinok1996Rank}.

\begin{figure}[!hb]
   \centering
   \includegraphics[scale=0.2]{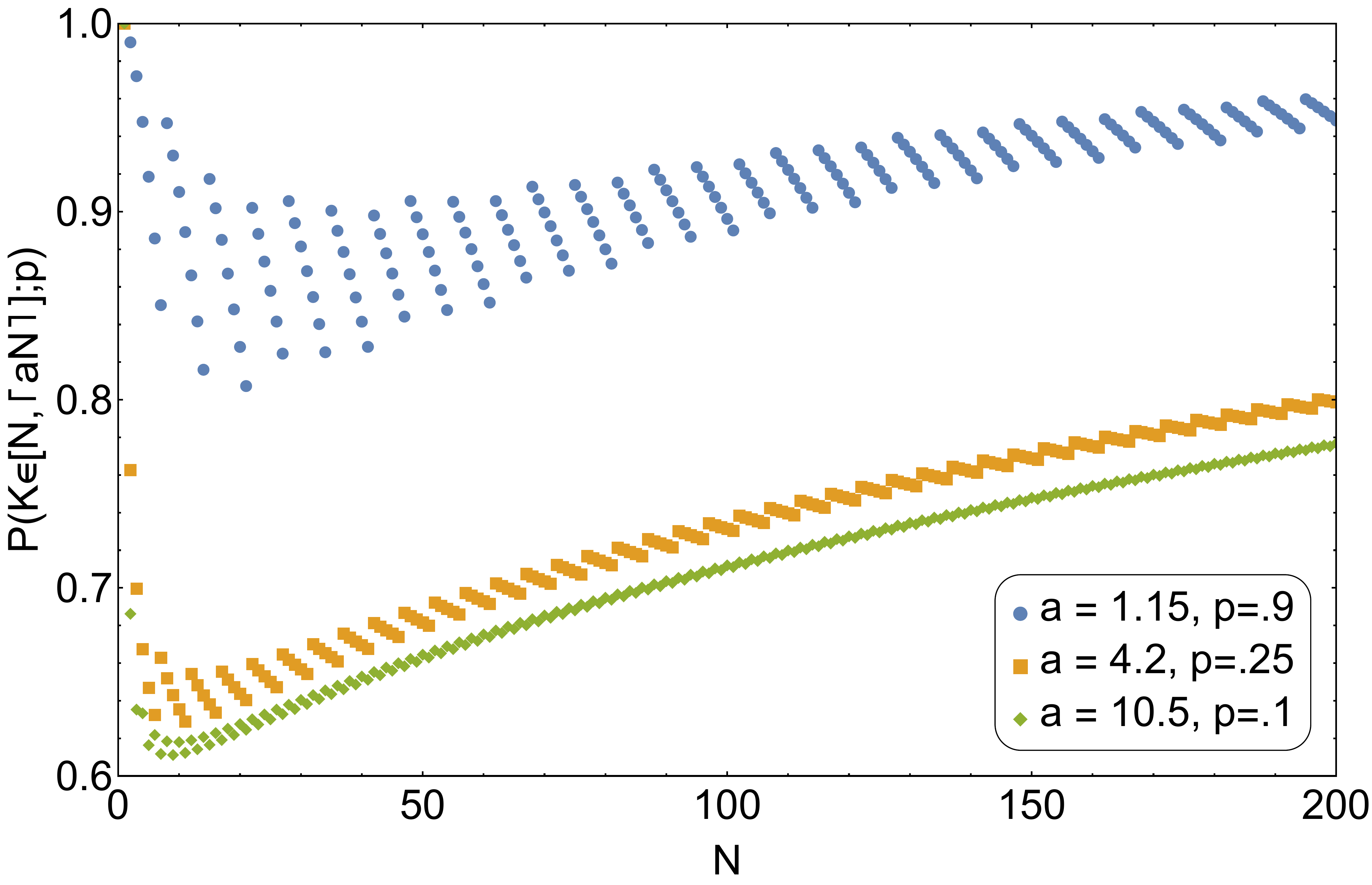}
\caption{Success probability in Eq. (\ref{eq:probability}) of generating a sample in the RNBS scheme in Fig. \ref{setup}  in the case of $\ceil{a N}$ sources emitting photons with probability $p$, for different values of $a$ and $p$ according to Eq. (\ref{eq:sourcenumber}) . Remarkably, for $p=.9$ already $\ceil{1.15 N}$ sources allow to achieve a success probability always larger than $.8$ and rapidly converging to $1$ for large values of $N$. In the other two depicted cases, the probability is always larger than .6 for  any value of $N$ even for the much smaller probability values $p = .1, .25$.}
	\label{fig:successprobability}
\end{figure}

\textit{Experimental scalability}.

We now demonstrate that RNBS allows an enhancement in the experimental scalability with respect to any other variants of boson sampling so far introduced by allowing to generate samples with a probability arbitrary close to one in absence of losses. 

The probability to successfully generate a sample at each experimental run is determined not only by the number $\ceil{aN}$ of input sources but also on the probability $p = \gamma^2$ for each source with squeezing parameter $\gamma$ to generate at least one photon. As mentioned before for small values of $\gamma$ one may be interested only in single photon emissions occurring with probability $p = (1-\gamma^2) \gamma^2$.
In particular, the total probability to experimentally generate any of the allowed samples ${\{ n_s ; n_d \}_{K}}$ for all possible random number  $K\in [N,\ceil  {a N}]$ of occupied input ports  can be written as \cite{Hartley1951_ChartIncompleteBetaFunction}
\begin{equation}
	\label{eq:probability}
P(K\in [N,\ceil  {a N}];p) = \sum_{K=N}^{\ceil{aN}}  \left(  \begin{array}{c} \ceil{aN} \\ K \end{array} \right)  p^{K} \big(1-p\big)^{\ceil{aN}-K}.
\end{equation}
Indeed $P$ is the cumulative probability for at least $N$ successful trials in a Bernoulli process with a total number of $\ceil{aN}$ trials and a success probability of $p$.
The corresponding binomial probability distribution for the number $K$ of successful trials has the mean value $\bar{K}=\ceil{aN} p$ and the standard deviation $\Delta K = \sqrt{\ceil{aN}} \sqrt{p(1-p)}$ \cite{Grinstead1998_Introductionprobability}.
Therefore, if 
\begin{equation}
	\label{eq:sourcenumber}
a > 1/p,
\end{equation}
 by considering all the values of $K$ in the interval  $[N,\ceil  {a N}]$ of length growing linearly with $N$, the cumulative probability $P(K\in [N,\ceil  {a N}];p) \rightarrow 1$ for large values of $N$. The exemplary cases  in Fig.~\ref{fig:successprobability} demonstrate the increasing of P for larger and larger values of $N$  converging toward an almost deterministic behavior.
Since the probability of postselecting only the successful experimental runs for $K\geq N$ approaches one for large values of $N$, the implementation of the RNBS problem with only a linear number of sources is practically deterministic.

We also emphasize that in the case of generalised RNBS for larger values of $p = \gamma^2$  the number ${\ceil{aN}}$ of required input sources becomes lower as expected according to Eq.  (\ref{eq:sourcenumber}) and as evident in Fig.~\ref{fig:successprobability}. 

\textit{RNBS with photon bunching events.}
Remarkably, RNBS schemes could be also developed  by allowing photon bunching occurrence at the detectors as long as at least $N$ detectors click. Indeed, also in this case the rank of the sampling matrices $U^{\{ n_s ; n_d \}_{K}}$, with values of $n_d$ higher than 1 for bunching events, is bounded from below by $N$  and the  approximate RNBS problem would be at least as hard as boson sampling and SBS. Furthermore, this may allow to reduce the total number of interferometric channels since it is not necessary any more to suppress bunching events according to the bosonic birthday paradox by requiring $M \sim N^2$ channels \cite{Aaronson2011_computationalcomplexitylinear}. In addition, allowing bunching events independently of the number $K' \geq N$ of detectors which click can further enhance the success probability of RNBS with respect to boson sampling and SBS with a fixed number $K'= N$ of clicking detectors.

\textit{Discussion.}
We have introduced in this letter the problem of Boson Sampling based on samples with a Random Number of input photons (RNBS). We have shown that this problem is at least as computationally hard as boson sampling and SBS either in the experimental scenarios where only single-photon emission are relevant or the one when multiple photon emissions from any of the sources can also be considered in the sampling process.  In both scenarios we have demonstrated that it is possible to generate a sample for large values of  $N$ with a probability which increases with N approaching to one by using  only probabilistic sources  linear in number.

Such technique outperforms any current boson sampling schemes where instead the sampling success probability decreases as the root of $N$. In addition, in the second RNBS scenario, by taking advantage of multiphoton emissions from sources with higher squeezing parameter, it is possible to  increase the  probability for each source to generate photons and consequently to decrease the required linear number of sources according to Eq. (\ref{eq:sourcenumber}) .

These results may also be extended to  Gaussian boson sampling experiments where it is possible to sample at the detector output over events where $N$ or more detectors click. Furthermore, it would be exciting to benchmark in future work RNBS against current classical algorithms to verify if any increase  in computational resources is needed with respect to the classical simulation of boson sampling and SBS, providing further insight in the boson sampling problem size needed to establish quantum computational supremacy.

In conclusion, this work can pave the way to scalable experimental demonstration of quantum computational supremacy and open further exciting questions toward a deeper understanding of  the computational hardness of multiphoton interference.

%

\begin{acknowledgments}
The authors are grateful to S. Aaronson, A. Arkhipov and P. Facchi for discussions
related to this work. V. T. is also thankful to W. P.
Schleich for the time until the summer of 2016 passed at the Institute of Quantum
Physics in Ulm where the ideas behind this work
started to flourish.
V.T. is grateful to Danilo Triggiani for his help in formatting the figures in the paper.
\end{acknowledgments}

\bibliography{RNBS}{}
\bibliographystyle{ieeetr}
\bibliographystyle{apsrev4-1}
\end{document}